\documentclass[apj]{emulateapj}
\usepackage{color}

\newcommand{\tn}[1]{{\rm #1}}

\begin{document}

\title{Testing Gravity with Quasi-Periodic Oscillations from accreting Black Holes:\\ the Case of the Einstein-Dilaton-Gauss-Bonnet Theory }

\author{Andrea Maselli\altaffilmark{1,4}, Leonardo Gualtieri\altaffilmark{1}, Paolo Pani\altaffilmark{2,1}, Luigi Stella\altaffilmark{3}, and Valeria Ferrari \altaffilmark{1}}
\altaffiltext{1}{Dipartimento di Fisica, Universit\`a di Roma ``La Sapienza'' \&   Sezione INFN Roma1, P. A.Moro 5, 00185, Roma, Italy.}
\altaffiltext{2}{CENTRA, Departamento de F\'{\i}sica, Instituto Superior T\'ecnico,  Universidade de Lisboa, Avenida~Rovisco Pais 1, 1049 Lisboa, Portugal.}
\altaffiltext{3}{INAF-Osservatorio Astronomico di Roma, via di Frascati 33, 00040, Monteporzio Catone, Roma, Italy.}
\altaffiltext{4}{Center for Relativistic Astrophysics, School of Physics, Georgia Institute of Technology, Atlanta, GA 30332, USA.}

\begin{abstract}
Quasi-Periodic oscillations (QPOs) observed in the X-ray flux emitted by accreting black holes are associated with phenomena occurring near the horizon.  Future very large area X-ray instruments 
will be able to measure QPO frequencies with very high precision, thus probing this strong-field region. Using the relativistic precession model, we show the
way in which QPO frequencies could be used to test general relativity (GR) against those alternative theories of gravity
which predict deviations from the classical theory in the strong-field and high-curvature regimes. We consider one of the best-motivated
high-curvature corrections to GR, namely the Einstein-Dilaton-Gauss-Bonnet theory, and show that a
detection of QPOs with the expected sensitivity of the proposed ESA M-class mission {\it LOFT} would set the most stringent
constraints on the parameter space of this theory.
\end{abstract}

\section{Introduction}\label{intro}

Since its formulation a century ago, general relativity (GR) has
successfully passed a number of observational and experimental tests
\citep{Will:2014xja}.  However, since these tests mostly probed the
weak-field regime of gravity, a number of strong-field/high-curvature GR predictions
still remain to be verified \citep{Psaltis:2008bb,Yunes:2013dva}.

Astrophysical systems involving collapsed objects,
i.e. black holes (BHs) and neutron stars (NSs), provide us with ways to
study the properties of the strong gravitational fields in their close
surroundings.
Most of the radiation emitted by matter accreting onto BHs and NSs
originates very deep in the gravitational fields of these objects, at
distances down to a few gravitational radii $r_g = GM/c^2$.  The motion of this
matter holds the potential to study the properties of gravity in the
strong-field regime and to verify some of the yet untested and key
predictions of GR. X-ray observations of matter
accretion onto stellar mass BHs and NSs in binary systems and onto
supermassive BHs in active galactic nuclei (AGNs) have singled out a few
powerful diagnostics over the last two decades. Spectral diagnostics
include the soft X-ray continuum emission of accretion disks, whose
highest temperature has been used to measure the innermost disk radius,
likely extending to the innermost stable circular orbit (ISCO), and
thus to infer the spin of stellar mass BHs (for a review see
\citet{McClintock:2011}). The extremely broad and redshifted iron
K$\alpha$ line around $\sim 6$~keV, observed in X-ray binaries as well as
AGNs, has been extensively exploited for the same purpose;  it also
carries key information about the dynamics, emissivity, and geometrical
properties of the inner disk regions (for a recent review see
\citet{Fabian:2013}).
Tests of GR based on observations of the soft X-ray continuum emission and of the iron K$\alpha$ line in BH accretion disks, have been discussed e.g. in \citet{Bambi:2011jq,Johannsen:2012ng,Bambi:2013sha}. 

Very fast flux variability produced by matter orbiting  close to BHs and NSs 
has long been considered a potential probe of geodetic motion in the strong-field regime 
(e.g. \citet{Sunyaev:1972}). Moreover, for stellar-mass BHs it also probes the high-curvature regime of the theory.

This diagnostic only began to come to fruition in the late 90s, when 
fast quasi-periodic oscillations (QPOs) at X-ray energies and with frequencies close 
to those expected from bound orbits at characteristic radii of $\lesssim
10\ r_g$, were discovered. 
Different QPO modes excited at the same time were studied in a number of NS X-ray 
binaries, whose variable frequency extended to $> 1$~kHz in the highest mode. A similar
QPO phenomenology was observed in several BH X-ray binaries, where frequencies 
up to $\sim 450$~Hz were detected (for a review see \citet{vanderklis:2006}). In a few 
cases QPOs were detected also in the X-ray flux of supermassive AGNs ( e.g. \citep{Gierlinski:2008}). 
Several models have been proposed to interpret the QPO phenomenon, virtually all of 
them involving the fundamental frequencies of motion of matter in the strong-field regime;
these are drastically different from their Newtonian equivalent, such that no weak field 
expansion of GR can approximate them (for a recent review see \citet{Belloni_Stella:2014} 
and references therein). 

The QPO signals can provide a powerful diagnostic of strong
gravitational fields and for regions with high curvature; BHs are especially 
promising in this respect, by virtue of their simplicity. First, the so-called 
{\it no-hair theorems} (see e.g.
\citet{Hawking:1973uf} and references therein) demonstrated that the
spacetime of a stationary BH in GR depends only on its mass and angular
momentum. Second, unlike NSs, BHs do not possess stable magnetic
fields, or a "hard" surface or boundary layer that can alter
considerably the dynamics of matter inflow at the smallest radii.
Simultaneous QPO modes in BHs have so far been detected only in a
few cases; moreover, the accuracy of the corresponding  
frequency measurements is limited to $> 1-2$\%.  
Despite these limitations, the application of the
Relativistic Precession Model (RPM, see Section 2.1) to the QPOs from
the BH X-ray binary GRO J1655-40 \citep{Motta:2013wga}, the only BH
binary system in which three simultaneous QPOs were observed, yielded
precise measurements of the BH mass and spin, the former being in
full agreement with the mass derived from optical observations.

With the development of very high throughput X-ray instrumentation, it
is expected that  simultaneous QPO signals will be detected in a variety
of BHs and their frequency will be measured to high precision and accuracy, such
that quantitative tests of GR predictions  in the strong-field/high-curvature
regime will become feasible. X-ray astronomy satellites that can
achieve these goals have been actively studied in recent years. In
particular, the proposed ESA X-ray satellite {\it LOFT}, with its extremely
high effective area (up to $\sim 10$~m$^2$) and good CCD-type spectral
resolution ($\sim 180$~eV) in the classical X-ray range 
(2 - 50~keV), offers the best prospects for exploiting the
QPO diagnostic \citep{Feroci_et_al:2012}.

As with other astrophysical systems (e.g. the pulse timing of
relativistic binary pulsars \citep{Kramer:2014} and extreme mass-ratio
inspirals \citep{Amaro:2014}), QPOs provide in principle two different
methods to test GR. In the first method, GR predictions  are
directly compared to measurements and tested for consistency.  In the
second method, alternative theories are introduced, and their parameters
are constrained to demonstrate whether GR is confirmed as the best theory of
gravity. 

Some degree of ``redundancy" is required in order to apply the second
method, in that the measurements must be sufficient to constrain (or infer)
more parameters than those required by GR alone. In this context,
alternative gravity theories have been introduced by adopting two
different approaches \citep{Psaltis:2009xf}. In a bottom-up approach,
one parametrizes the BH spacetime in a phenomenological way and, once a
deviation from GR is found (or constrained) in terms of these
parameters, one tries to interpret such a deviation (or constraint) in
terms of an alternative theory.
However, the parametrizations that have been proposed in the literature up to now 
either do not reproduce BH solutions of known theories alternative to GR
\citep{Collins:2004ex,Glampedakis:2005cf,Johannsen:2010xs}
or are very involved, and therefore are to some extent
impractical \citep{Vigeland:2011ji}.
In a top-down approach, instead, one
considers modifications of GR, possibly inspired by fundamental physics
considerations, and then works out the predictions of such
modifications to be tested against observation.  We adopt here a
top-down approach to calculate the modified fundamental frequencies of
motion that underlay QPO models.
We mention that bottom-up approaches have been employed to test GR
using QPO signals from accreting BHs, in
\citet{Johannsen:2010bi,Bambi:2012pa,Bambi:2013fea}.

Among the gravity theories which have been proposed as alternatives to GR (see e.g. the reviews of
\citet{Psaltis:2008bb}, \citet{Yunes:2013dva}, \citet{Will:2014xja}) we select the Einstein-Dilaton-Gauss-Bonnet (EDGB) theory \citep{Kanti:1995vq}, in which the Gauss-Bonnet invariant 
\begin{equation}
{\cal
R}^2_{\tn{GB}}=R_{\alpha\beta\delta\gamma}R^{\alpha\beta\delta\gamma}-4R_{\alpha\beta}R^{\alpha\beta}+R^2
\label{GB}
\end{equation} 
is included in the action, coupled with a scalar field. The reasons for this choice are as follow.
\begin{itemize}
\item The most natural way to modify the strong-field/high-curvature regime of gravity is 
to include in the action a quadratic term in the curvature tensor.
\item If the equations of motion have third- (or higher)  order
derivatives, then the theory is subject to Ostrogradsky's
  instability \citep{Woodard:2006nt}. The Gauss-Bonnet invariant is the only quadratic term in the curvature which leads
  to second-order field equations, thus avoiding this instability. Theories with quadratic curvature invariants different
  from~(\ref{GB}) should be treated as effective theories, in which higher-order terms are assumed to be present in the
  action and neglected in some regime.
\item 
Since ${\cal R}^2_{\tn{GB}}$ is a total derivative, it would not
contribute to the field equations unless it is coupled
to a scalar field, as in the EDGB theory.
\item The Gauss-Bonnet term can be seen as the first term in an
expansion including all possible curvature invariants and their powers, as suggested by
low-energy effective string theories (see e.g. \citet{Moura:2006pz} and
references therein). 
\item Scalar-tensor theories which do not include quadratic (or higher) curvature
invariants (see e.g.  \citet{fujii2003scalar} and references therein) do
not introduce  strong-field/high curvature corrections to GR\footnote{Except - in the case 
of neutron stars - for specific solutions, for instance, such as those discussed in \citet{Damour:1993hw}.}. Moreover,
stationary BHs in these theories satisfy the no-hair theorems of GR
\citep{Sotiriou:2011dz}. It is then impossible to test GR against
these theories using BHs close to stationarity. Similar results
apply to $f(R)$ theories. In the EDGB theory, instead, BH solutions are
different from those of GR
\citep{Mignemi:1992nt,Kanti:1995vq,Pani:2009wy,Yunes:2011we,Sotiriou:2013qea}. Testing these
differences is the goal of this work.  
\end{itemize}

In this paper, we calculate the azimuthal and epicyclic frequencies of a
slowly rotating BH in EDGB gravity and find that these differ from
their GR equivalent by up to $\sim4\%$. A similar computation has been
carried out for dynamical Chern-Simons gravity (a different theory with
quadratic curvature terms, see \citet{Alexander:2009tp} for a review)
in \citet{Vincent:2013uea}, finding that
deviations from GR predicted by that theory are much smaller. 
Using the RPM, we show that the differences between the
QPO frequencies predicted by GR and EDGB gravity, while undetectable
with currently available BH QPO measurements, can be large enough to be
measured with the next generation of large area X-ray instruments. This paper is 
organized as follows. In Section \ref{sec:QPO} we summarize the
RPM and the procedure for computing the azimuthal and epicyclic frequencies
in a BH spacetime. In Section \ref{sec:EDGB}, we briefly discuss EDGB
gravity, the solutions of this theory describing slowly rotating BHs,
the geodesics in this spacetime, and the approach to compute the
epicyclic frequencies.  In Section \ref{sec:results} we compute the QPO
frequencies in EDGB gravity and show how they can be used to test GR
against EDGB theory. Finally, in Section \ref{sec:concl} we draw the conclusions.

\section{Quasi-periodic oscillations in the X-ray flux of accreting BHS}\label{sec:QPO}
Most QPO models were originally proposed in the context of 
accreting NSs, with the faster of the twin kHz QPO modes 
often being interpreted as arising from the azimuthal
motion of matter in the inner disk region
(for a review see \citet{vanderklis:2000}).
When BH QPOs were discovered, it was realized that despite
their relatively low amplitudes and simpler phenomenology, 
they provided a ``cleaner'' environment to study the 
properties of strong gravitational fields and of high-curvature regions, through the 
motion of matter that generates the QPOs.

Building on the similarity between the QPO modes that are 
observed in NSs and BHs, especially the presence of 
a QPO pair at higher frequencies, a few models  
were applied to both types of systems.
Especially successful among these are the RPM, 
upon which we base our analysis on here, 
and the epicyclic resonance model (\citet{Kluzniak_Abramowicz:2001}, 
\citet{Abramowicz_Kluzniak:2001}) which will be discussed in a future
work.

\subsection{The relativistic precession model}\label{sec:RPM}

The RPM was formulated in order to  
interpret the twin QPOs around $\sim 1$~kHz as well as a low-frequency
QPO mode (the so-called Horizontal Branch Oscillations, HBOs) 
of NSs in low-mass X-ray binaries
(\citet{Stella_Vietri:1998}, \citet{Stella:1998mq}).
The higher- and lower-frequency kHz QPOs are identified
with the azimuthal frequency $\nu_\varphi$, and the periastron
precession frequency, $\nu_\tn{per} = \nu_\varphi-\nu_r$, of matter orbiting
in quasi-circular orbits 
at a given radius $r$ of the innermost disk region, with $\nu_r$ being
the radial epicyclic frequency. 
The low-frequency QPO mode  is instead related to the nodal 
precession frequency, $\nu_\tn{nod}=\nu_\varphi - \nu_\theta$, where 
$\nu_\theta$ is the vertical epicyclic frequency. 
$\nu_\tn{nod}$  is emitted at the same radius 
where the signals at $\nu_\varphi$ and $\nu_\tn{per}$ are produced.
Correlated QPO frequency variations that are observed 
from individual NSs at different times are well reproduced in 
the RPM as a result of variations in the radius at which the QPOs 
are emitted.

The RPM was first applied to BH systems by \citet{Stella_V_M:1999}. 
A complete application
to BHs involving three QPO modes became possible only recently,
following the realization that in GRO J1655-40 the 
BH equivalent of the HBOs, the so-called type C low-frequency QPOs
\citep{2005ApJ...629..403C}, in one instance were detected simultaneously 
with two high-frequency QPOs (see 
\citet{Motta:2013wga} and references therein).
The centroid frequency of these QPOs 
was measured by the PCA on board the {\it Rossi X-ray Timing 
Explorer (RXTE)} with 
1$\sigma$ uncertainties in the 1-2\% range. 
Their values are
\begin{equation}
\label{freq}
\nu_\varphi= 441^{+2}_{-2}~\tn{Hz},\ 
\nu_\tn{per}= 298^{+4}_{-4}~\tn{Hz},\ 
\nu_\tn{nod}= 17.3^{+0.1}_{-0.1}~\tn{Hz}~,
\end{equation}
where the $\nu$ symbols refer to the RPM interpretation.
By fitting the three simultaneous QPOs (hereafter the QPO triplet) 
with the RPM 
frequencies from the Kerr metric,
precise values of
$M=(5.31\pm 0.07) M_\odot$, 
$a^\star=J/M^2 = 0.290 \pm 0.003$, and 
radius\footnote{$M$, $J$ are the Arnowitt-Deser-Misner BH mass and angular momentum,
respectively.} from which 
the QPO triplet originated were obtained through the sole use of X-ray timing.

The former value is fully consistent with the BH mass as inferred from 
optical/NIR spectro-photometric observations, $(5.4\, \pm 0.3) M_\odot$ 
\citep{Beer_Podsiadlowski_2002}. 
The QPO radius  was found 
to be $(5.68 \,\pm 0.04)\, r_{g}$, i.e. only $\sim 13$\% larger than $R_\tn{ISCO}$, 
very deep in the gravitational field of the BH\footnote{ For the values above, the BH 
event horizon is at $ 1.96\, M$.}. 

The detection of a single QPO triplet 
yields only the above-mentioned three quantities with no 
redundancy. The relative width of the three power spectrum QPO peaks 
in GRO J1655-40 
was found to be consistent with being due to jitter variations of the 
radius where QPOs are generated, suggesting that future, high-throughput
X-ray observations can reveal frequency variations of the 
QPO triplet as a function of time in accreting BH systems.
If more triplets are detected, then additional information concerning the properties
of strong-field/high-curvature gravity very close to BHs can be derived:
for instance, it would be possible to study the radial dependence of the fundamental
frequencies. 
Since the signal-to-noise ratio of incoherent power spectrum signals 
(such as QPOs) scales linearly with the rates in photon 
counting instruments, larger area X-ray detectors of the future will 
provide a correspondingly higher precision in QPO measurement. 
Our calculations here are based on the {\it LOFT}-LAD instrument 
which, due to its effective area, will provide a factor of-$\sim 15$ 
improved precision relative to the RXTE-PCA values quoted 
above (see Eq.~(\ref{freq})).

\subsection{The epicyclic frequencies of a rotating black hole}\label{sec:epyc}
Here, we derive the expressions for the epicyclic frequencies (see also
\citet{wald2010general,1999MNRAS.304..155M,Abramowicz:2002iu,Vincent:2013uea}).  In this
derivation, we only require that the spacetime is stationary,
axisymmetric, circular (i.e. invariant to the simultaneous inversion of
time and of the azimuthal angle $\varphi$), and symmetric across the
equatorial plane. Circularity and equatorial symmetry are satisfied by
(stationary, axisymmetric) BH solutions in GR and by most BH solutions
in alternative theories of gravity; in particular, they are satisfied by
(stationary, axisymmetric) BH solutions in EDGB  gravity.

The metric of a stationary, axially symmetric spacetime which  satisfies the
circularity condition can be expressed in the form (see e.g.
\citet{chandrasekhar1983mathematical})

\begin{equation}
ds^2=g_{tt}dt^2+g_{rr}dr^2+g_{\theta\theta}d\theta^2+2g_{t\varphi}dtd\varphi +g_{\varphi\varphi}d\varphi^2\,,\label{genmetric}
\end{equation}
where $g_{\mu\nu}=g_{\mu\nu}(r,\theta)$.  Let $u^\mu={\dot
x}^\mu=dx^\mu/d\tau$ be the four-velocity of a massive particle ($\tau$
being the proper time). Stationarity and axisymmetry imply that 
there are two constants of geodesic motion, $E=-u_t$ and $L=u_\varphi$, which we
refer to as  the energy and  the angular momentum per unit mass of the
moving particle.
We also define the proper angular momentum
$l=L/E$, and the potential 
\begin{equation} {\cal
U}(r,\theta)=g^{tt}-2lg^{t\varphi}+l^2g^{\varphi\varphi}\,.\label{defU}
\end{equation}
Since $g^{\mu\nu}u_\mu u_\nu=-1$, 
\begin{equation} g_{rr}{\dot
r}^2+g_{\theta\theta}{\dot\theta}^2+E^2{\cal U}(r,\theta)=-1\,.
\label{normaliz} 
\end{equation} 
If we consider equatorial motion, then Eq.~(\ref{normaliz}) gives
 ${\dot r}^2=V(r)$, where
$V(r)=-g_{rr}^{-1}E^2\left[E^2{\cal U}(r,\pi/2)+1\right]$ is an effective
potential.

For a circular, equatorial orbit at $r=\bar r$, ${\dot r}=0$ and, from
Eq.~(\ref{normaliz}), we get ${\cal U}_{,r}(\bar r,\pi/2)={\cal
U}_{,\theta}(\bar r,\pi/2)=0$.  $E$ and $L$ can be determined by
imposing $V(\bar r)=V'(\bar r)=0$, where the prime indicates
differentiation with respect to $r$ and  the further condition $V''=0$ yields
the ISCO radius. The four-velocity on a
circular, equatorial orbit has the form $u^\mu=u^0(1,0,0,\Omega)$, where
the angular velocity $\Omega=2\pi\nu_\varphi$ can be found by solving
the algebraic equation \begin{equation} g_{tt,r}+2\Omega
g_{t\varphi,r}+\Omega^2g_{\varphi\varphi,r}=0\,.\label{wphi}
\end{equation}

Let us now consider a small perturbation of a circular, equatorial orbit, i.e.,
\begin{equation}
r(t)=\bar r+\delta r(t)\qquad \qquad \theta(t)=\frac{\pi}{2}+\delta\theta(t)
\end{equation}
with $\delta r\sim e^{2\pi i\nu_rt}$, $\delta\theta\sim e^{2\pi
i\nu_\theta t}$, and $\delta r\ll\bar r$, $\delta\theta\ll\pi/2$.
$\nu_\varphi$, $\nu_r$, and $\nu_\theta$ are the azimuthal and the epicyclic frequencies.
Eq.~(\ref{normaliz}) yields 
\begin{eqnarray}
&&-g_{rr}(u^02\pi\nu_r\delta
r)^2-g_{\theta\theta}(u^02\pi\nu_\theta\delta\theta)^2+E^2\left({\cal
U}\left(\bar r,\frac{\pi}{2}\right)\right.\nonumber\\
&&\left.+\frac{1}{2}\frac{\partial^2{\cal U}}{\partial r^2} \left(\bar
r,\frac{\pi}{2}\right)\delta r^2+ \frac{1}{2}\frac{\partial^2{\cal
U}}{\partial\theta^2} \left(\bar
r,\frac{\pi}{2}\right)\delta\theta^2\right)=0\,.\label{pass}
\end{eqnarray} 
Since
$E=-(g_{tt}+\Omega g_{t\varphi})u^0$, Eq.~(\ref{pass}) implies
\begin{eqnarray} \nu_r^2&=&\frac{(g_{tt}+\Omega
g_{t\varphi})^2}{2(2\pi)^2g_{rr}}\frac{\partial^2{\cal U}}{\partial
r^2}\left(\bar r,\frac{\pi}{2}\right)\,,\nonumber\\
\nu_\theta^2&=&\frac{(g_{tt}+\Omega
g_{t\varphi})^2}{2(2\pi)^2g_{\theta\theta}}\frac{\partial^2{\cal
U}}{\partial\theta^2} \left(\bar r,\frac{\pi}{2}\right)\,.\label{epic}
\end{eqnarray}

\section{Spinning BHs in Einstein-Dilaton-Gauss-Bonnet gravity}\label{sec:EDGB}
In this Section, we briefly describe the EDGB theory and its slowly rotating
BH solution.
\subsection{EDGB gravity}
EDGB theory is defined by the following action
\citep{Kanti:1995vq}~\footnote{Note that we use the same signature as in
\citet{Kanti:1995vq}, but the opposite sign convention for the
definition of the Riemann tensor.} 
\begin{equation}\label{EDGB}
S=\frac{1}{2}\int
d^{4}x\sqrt{-g}\left(R-\frac{1}{2}\partial_{\mu}\phi\partial^{\mu}\phi
+\frac{\alpha e^\phi}{4}{\cal R}^2_{\tn{GB}}\right)\ ,
\end{equation}
where the Gauss-Bonnet invariant ${\cal R}^2_{\tn{GB}}$ is defined as in
Eq.~(\ref{GB}) and $\alpha>0$ is a coupling constant (in the framework
of string theory, $\alpha$ corresponds to the Regge slope; the gauge
coupling constant has been fixed to $1$). We use units in which $G=c=1$;
with this choice, the scalar field $\phi$ is dimensionless and $\alpha$
has the dimension of a squared length.

The  field equations of the EDGB theory are found
varying the action (\ref{EDGB}) with respect to $g_{\mu\nu}$ and $\phi$:
\begin{eqnarray}
&&G_{\mu\nu}=\frac{1}{2}\partial_{\mu}\phi\partial_{\nu}\phi-
\frac{1}{4}g_{\mu\nu}\partial_{\alpha}\phi\partial^{\alpha}\phi-\alpha{\cal K}_{\mu\nu}\,,\label{EinsEq}\\
&&\frac{1}{\sqrt{g}}\partial_{\mu}(\sqrt{g}\partial^{\mu}\phi)=-\frac{\alpha}{4}e^{\phi}
{\cal R}^2_{\tn{GB}}\,, \label{KGeq}
\end{eqnarray}
where $G_{\mu\nu}=R_{\mu\nu}-\frac{1}{2}g_{\mu\nu}R$ is the Einstein tensor,
\begin{equation}
{\cal K}_{\mu\nu}=\frac{g_{\mu\rho}g_{\nu\lambda} +g_{\mu\lambda}g_{\nu\rho}}{8}\epsilon^{k\lambda\alpha\beta}
\nabla_{\gamma}\left(\epsilon^{\rho\gamma\mu\nu}R_{\mu\nu\alpha\beta} \partial_{k}e^\phi\right)\,,
\end{equation}
and $\epsilon^{0123}=-(-g)^{-1/2}$.

The solution of Equations (\ref{EinsEq}) and (\ref{KGeq}) describing a spherically symmetric BH 
was found by solving the
field equations numerically in \citet{Kanti:1995vq}. An analytical solution has been derived to second order in
$\alpha/M^2$ in \cite{Mignemi:1992nt,Yunes:2011we}.

\subsection{Slowly rotating black holes}\label{sec:slBH}
The solution of the EDGB equations (\ref{EinsEq}) and (\ref{KGeq})
which describes a slowly rotating BH was derived in \citet{Pani:2009wy}.
It was obtained as a perturbation of the spherically symmetric,
non-rotating BH solution derived in \citet{Kanti:1995vq}, to first
order in the BH spin ${a^\star}=J/M^2$.
This solution (as in the non-rotating case) only exists if
\begin{equation}\label{BHcond}
e^{\phi_{\tn{h}}}\leq \frac{r^2_{\tn{h}}}{\alpha\sqrt{6}}\,, 
\end{equation}
where $r_{\tn{h}}$ and $\phi_{h}$ are the radial coordinate and the
scalar field evaluated at the horizon. We refer the reader to the
Appendix of \citet{Kanti:1995vq} for an explicit derivation of the
equations of motion of the spherically symmetric solution. From the
asymptotic behavior of the metric, we can extract $M$ and the dilatonic
charge $D$.  Since the field equations are invariant under the rescaling
$\phi\rightarrow\phi+\phi_0$, $r\rightarrow re^{\phi_0/2}$
(or equivalently $M\rightarrow Me^{\phi_0/2}$, $D\rightarrow De^{\phi_0/2}$), for each
value of $M$ there is only one solution to the non-rotating BH; the
dilatonic charge can be determined in terms of the BH mass.

As noted in \citet{Pani:2009wy}, by imposing asymptotic 
flatness (i.e. $\lim_{r\rightarrow\infty}\phi=0$), the condition
(\ref{BHcond}) can be written as
\begin{equation}\label{BHcond2}
0<\frac{\alpha}{M^2}\lesssim0.691\,.
\end{equation}
The best observational bound on the coupling parameter was derived
by~\citet{Yagi:2012gp} using the orbital decay rate of low-mass X-ray
binaries and reads\footnote{This corresponds to the bound
$\sqrt{\alpha}<1.9$ km derived in ~\citet{Yagi:2012gp} from observations
of the low-mass X-ray binary A0620-00. 
Note also that this bound was obtained by truncating the EDGB
theory to first order in the coupling, and does not necessarily hold in
the full theory that we are studying here. In order to convert our
notation and conventions to those of \citet{Yagi:2012gp}, we should
replace $\phi\rightarrow\sqrt{16\pi}\phi$ and $\alpha\rightarrow
16\sqrt{\pi}\alpha$.} $\alpha\lesssim 47M_\odot^2$, which is weaker than
the theoretical bound~(\ref{BHcond2}) for BHs with $M\lesssim 8.2
M_\odot$. Constraints of the same order of magnitude were obtained
in~\citet{Pani:2011xm} by studying slowly rotating NSs in the EDGB
theory, but they are limited by the uncertainties on the NS equation of
state. Thus, the range defined by the theoretical bound~(\ref{BHcond2})
is unconstrained to date (at least, for BHs with $M\lesssim 8.2
M_\odot$).

The spacetime metric of a slowly rotating BH at first order in the angular momentum 
can be written as
\begin{eqnarray}
ds^2&=&-f(r)dt^2+\frac{dr^2}{g(r)}+r^2(d\theta^2+\sin^2\theta d\varphi^2)\nonumber\\
&&-2r^2\sin^2\theta\omega(r)dtd\varphi\,.
\label{SRmetric}
\end{eqnarray}
Equations (\ref{EinsEq}) and (\ref{KGeq}) yield a set of differential
equations which allow us to compute the metric functions $f(r)$, $g(r)$,
$\omega(r)$ and the scalar profile $\phi(r)$.  The equations for $f(r)$,
$g(r)$ and $\phi(r)$, coincide with those of a non-rotating BH, which are discussed
in \citet{Kanti:1995vq}; the equation for $\omega(r)$ has the form
\begin{equation}\label{omegaeq}
\omega''(r)+\frac{G_{1}(r)}{G_{2}(r)}\omega'(r)=0\ ,
\end{equation}
where 
\begin{eqnarray}
G_{1}(r)&=&-\frac{r^2}{g}\left[\frac{f'}{f}-\frac{g'}{g}-\frac{8}{r}\right]
-\phi'\left[\frac{6}{r}-\frac{f'}{f}+\frac{3g'}{g}+2\phi'\right]\ +\nonumber \\
&&+2e^{\phi}r\phi''\ ,\\
G_{2}(r)&=&2r^2g^{-1}-2r e^{\phi}\phi'\ .
\end{eqnarray}
The solution of Eq.~(\ref{omegaeq}) can be written as
\begin{equation}
\omega(r)=C_{1}+C_{2}\int^r ds\ e^{-\int^sG_{2}/G_{3}dl}\, ,  
\end{equation}
where $C_{1}$ and $C_{2}$ are integration constants.
Since $f(r)$, $g(r)$, and $\phi(r)$ are known to be the result of 
numerical integrations, Eq.~(\ref{omegaeq}) has to be solved numerically as well.
  \footnote{In the small coupling limit, where $f(r)$, $g(r)$, and $\phi(r)$ are known analytically,
    Eq.~(\ref{omegaeq}) has been obtained in closed form in \citet{Pani:2011gy}. The analytic, small coupling solution
    has been extended to second order in the spin by \citet{Ayzenberg:2014aka}.}

Imposing the constraint that the spacetime is asymptotically flat, we obtain 
the asymptotic behavior
\begin{equation}\label{boundary}
\omega(r)\sim \frac{2J}{r^3}\,.
\end{equation}
For any given value of $J$, Eq.~(\ref{boundary}) and its first
derivative allow us to fix the integration constants $C_{1}$ and $C_{2}$.

The solution of (\ref{SRmetric}) has the form (\ref{genmetric}), with
$g_{tt}=-f(r)$, $g_{rr}=1/g(r)$, $g_{\theta\theta}=r^2$,
$g_{\varphi\varphi}=r^2\sin^2\theta$, and $g_{t\varphi}=-r^2\sin^2\theta\omega(r)$, 
and it is symmetric across the equatorial plane; therefore, the expressions of the azimuthal and
epicyclic frequencies $(\nu_\varphi$, $\nu_r$, and $\nu_\theta)$ presented
in Sec.~\ref{sec:RPM} also hold for the solution (\ref{SRmetric}).
Finally, as discussed in Sec.~\ref{sec:RPM}, we define the periastron
and nodal precession frequencies:
\begin{equation}
\nu_{\tn{nod}}=\nu_{\varphi}-\nu_{\theta}\quad \ ,\quad 
\nu_{\tn{per}}=\nu_{\varphi}-\nu_{r}\ .
\end{equation}

\section{Results}\label{sec:results}
\begin{table*}[ht]
\centering
\begin{tabular}{c|ccccc|ccccc|ccccc}
\hline
&\multicolumn{5}{c}{${a^\star}=0.02$}&
\multicolumn{5}{c}{${a^\star}=0.05$}&\multicolumn{5}{c}{${a^\star}=0.1$}\\
\hline
$\alpha/M^{2}$ & $\epsilon_{\varphi}\,\%$ & $\epsilon_r\,\%$ &
$\epsilon_{\theta}\,\%$ & $\epsilon_{\tn{nod}}\,\%$ &
$\epsilon_{\tn{per}}\,\%$
&$\epsilon_{\varphi}\,\%$ & $\epsilon_{r}\,\%$ &
$\epsilon_{\theta}\,\%$ & $\epsilon_{\tn{nod}}\,\%$ &
$\epsilon_{\tn{per}}\,\%$
&$\epsilon_{\varphi}\,\%$ & $\epsilon_{r}\,\%$ &
$\epsilon_{\theta}\,\%$ & $\epsilon_{\tn{nod}}\,\%$ &
$\epsilon_{\tn{per}}\,\%$
 \\
\hline
 0.691 & 2.66 & 1.07 & 2.65 & 7.4 & 3.34 & 2.85 & 1.16 & 2.82 & 9.5 & 3.58 & 3.26 & 1.37 & 3.16 & 13.0 & 4.06 \\
 0.576 & 1.30 & 0.55 & 1.30 & 4.6 & 1.63 & 1.43 & 0.62 & 1.40 & 6.7 & 1.78 & 1.73 & 0.80 & 1.64 & 10.2 & 2.12 \\
 0.418 & 0.58 & 0.26 & 0.58 & 3.2 & 0.72 & 0.67 & 0.32 & 0.65 & 5.2 & 0.82 & 0.91 & 0.50 & 0.83 & 8.6 & 1.09 \\
 0.237 & 0.18 & 0.07 & 0.17 & 2.3 & 0.22 & 0.26 & 0.11 & 0.23 & 4.3 & 0.32 & 0.47 & 0.23 & 0.34 & 7.7 & 0.57 \\
 0.104 & 0.05 & 0.03 & 0.04 & 1.8 & 0.05 & 0.11 & 0.08 & 0.09 & 3.9 & 0.12 & 0.31 & 0.23 & 0.23 & 7.2 & 0.34 \\
 0.0529 & 0.03 & 0.02 & 0.02 & 2.0 & 0.03 & 0.10 & 0.07 & 0.08 & 4.0 & 0.11 & 0.30 & 0.21 & 0.23 & 7.4 & 0.34 \\
 0.00298 & 0.02 & 0.00 & 0.01 & 1.8 & 0.02 & 0.08 & 0.04 & 0.06 & 3.8 & 0.10 & 0.27 & 0.16 & 0.20 & 7.2 & 0.32 \\\\
\hline
\hline
\end{tabular} \caption{Relative difference between the values
of the azimuthal/epicyclic frequencies derived in GR and in EDGB gravity,
computed
at the emission radius $r=1.1\,r_{\tn{ISCO}}$, for $M=5.3\,M_\odot$,
${a^\star}=(0.02,0.05,0.1)$, and different choices of $\alpha/M^2$.}
\label{tableresults} \end{table*}
\begin{figure*}[ht] \centering 
\includegraphics[width=5.9cm]{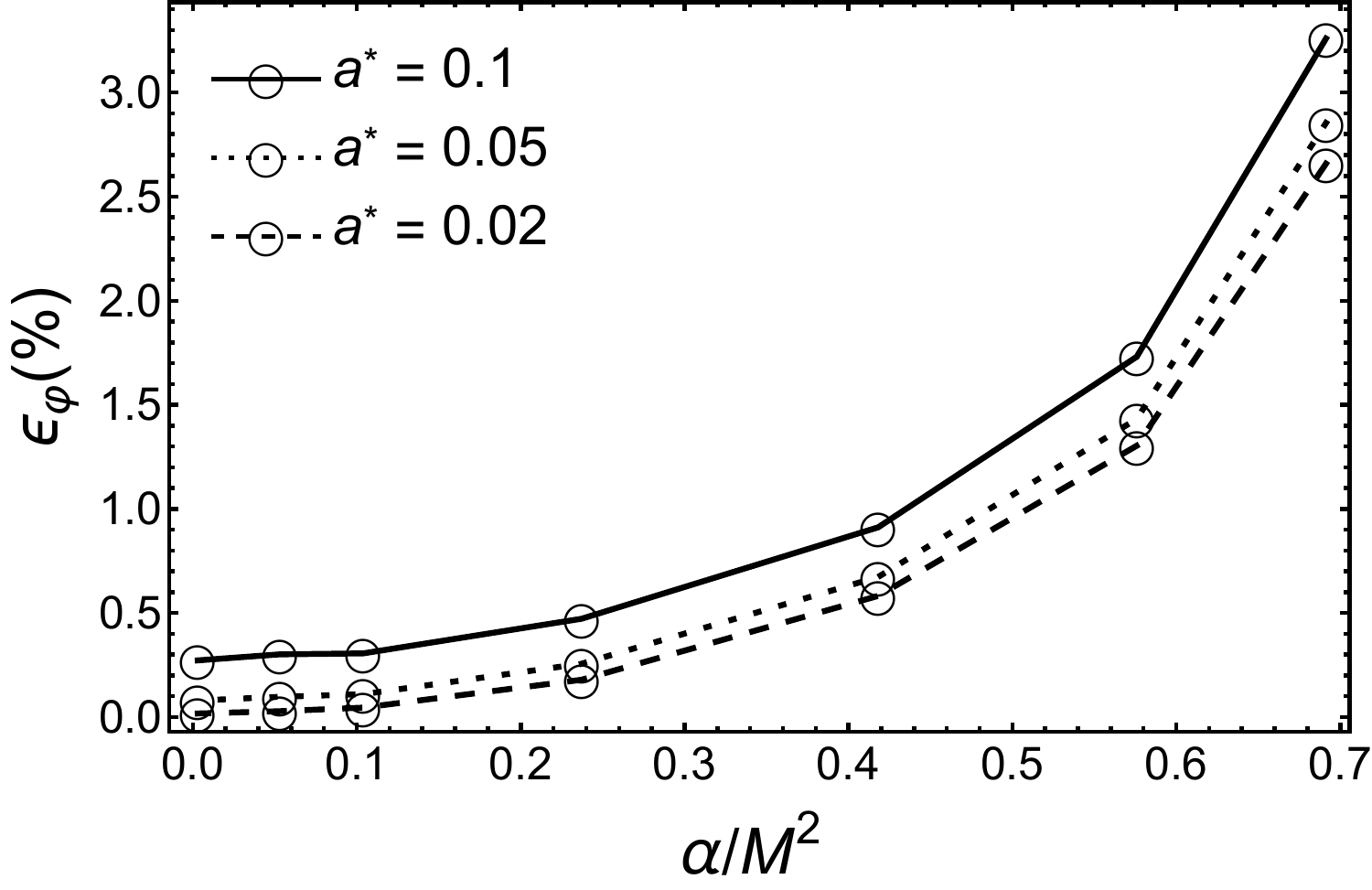}
\includegraphics[width=5.9cm]{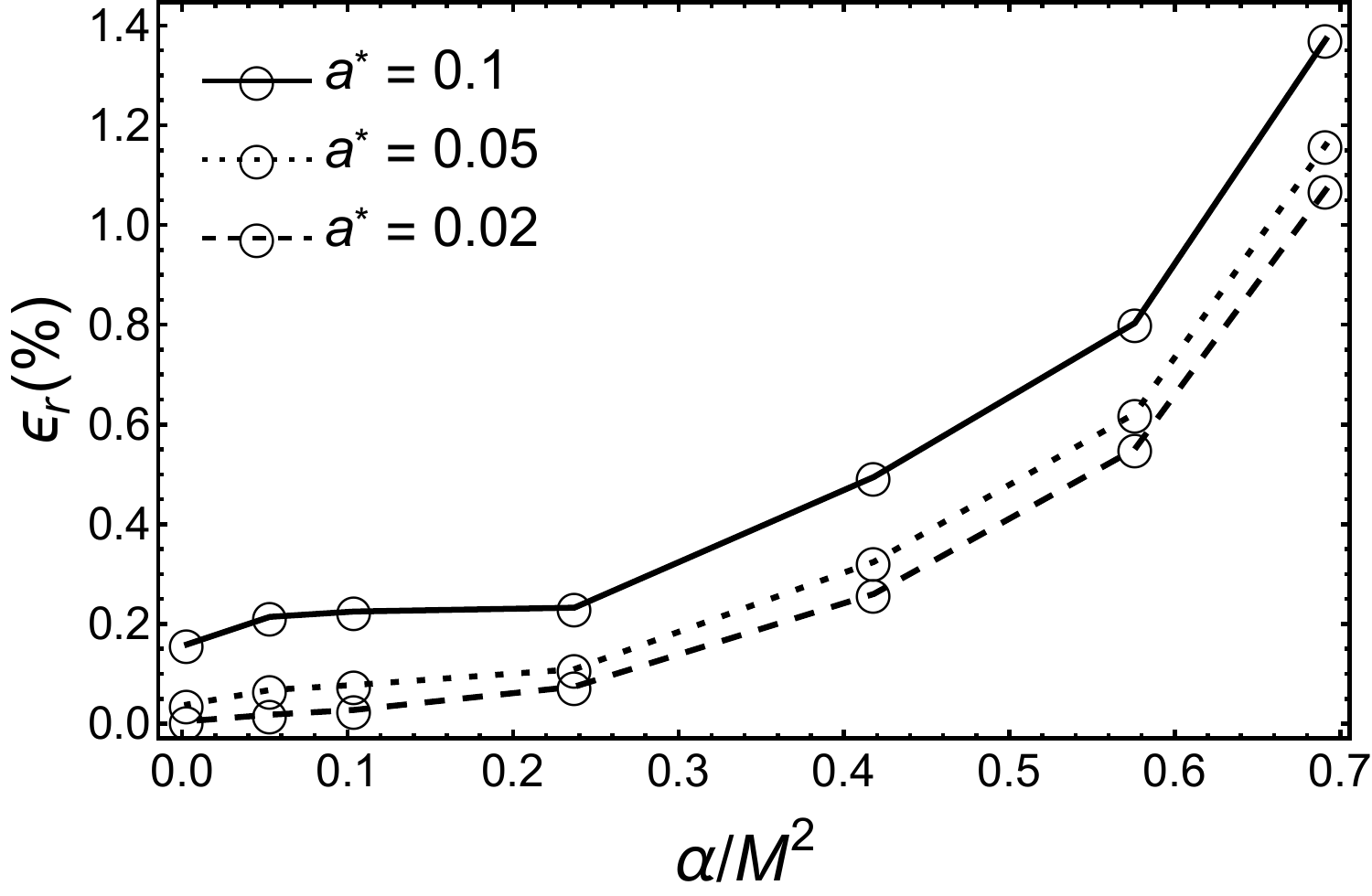}
\includegraphics[width=5.9cm]{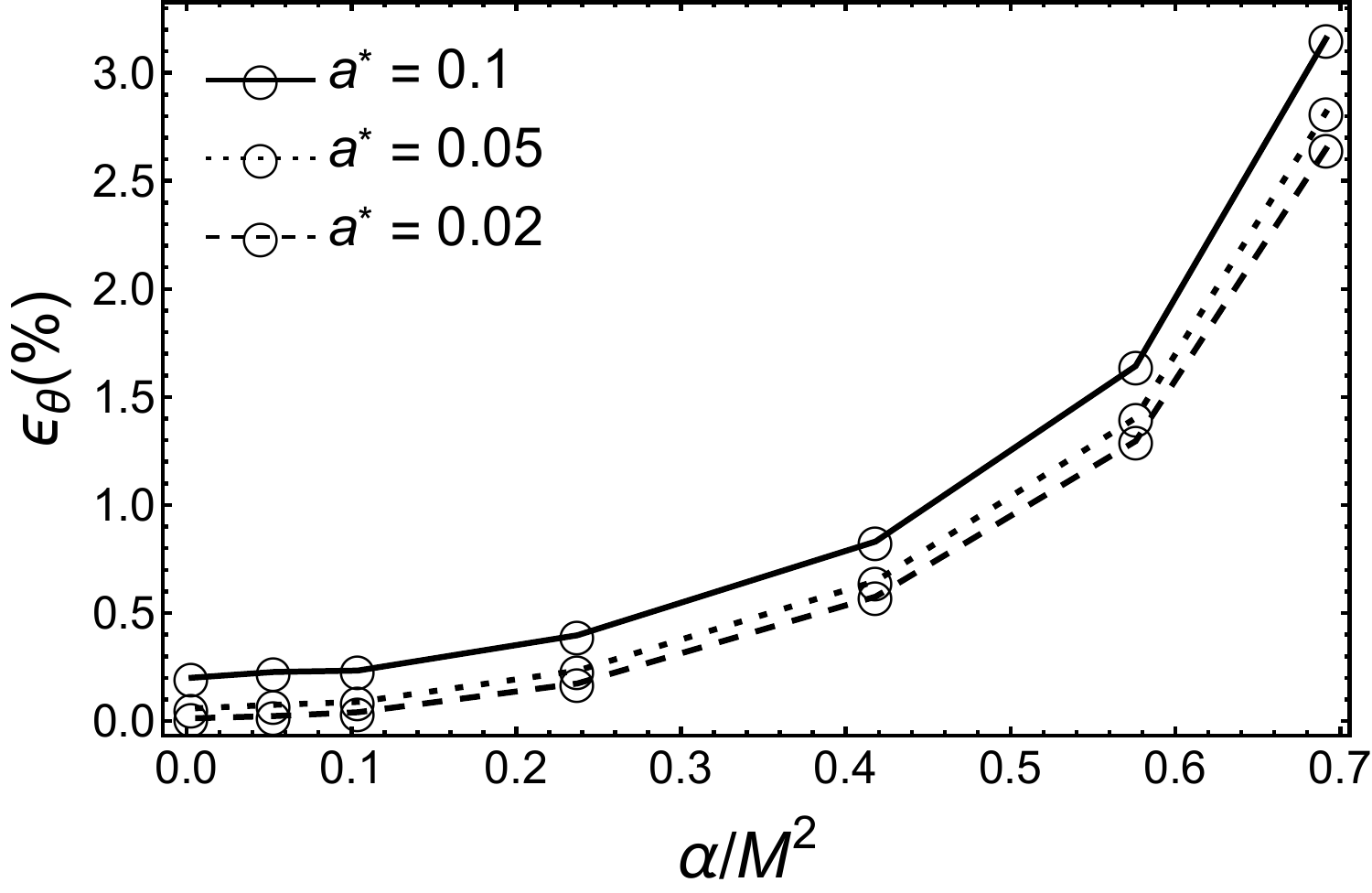} 
\caption{Relative
difference between the azimuthal/epicyclic frequencies
$(\nu_\varphi,\nu_r,\nu_\theta)$ computed in GR and in the EDGB theory 
plotted as a function of the coupling constant $\alpha/M^{2}$, for different
values of the BH spin parameter ${a^\star} =(0.02,0.05,0.1)$. All quantities
are computed at $r=1.1\,r_{\tn{ISCO}}$. 
} \label{discrepancy}
\end{figure*}
\subsection{Epicyclic frequencies in EDGB gravity}
We have computed the frequencies $\nu_\varphi$, $\nu_r$,
$\nu_\theta$, $\nu_{\tn{per}}$ and
$\nu_{\tn{nod}}$ of a slowly rotating BH in EDGB gravity as functions
of the BH mass and angular momentum, and  of
the coupling constant $\alpha/M^2$.
The solution discussed in Sec.~\ref{sec:slBH} describes rotating BHs to
first order in the spin ${a^\star}$. Therefore, our results are affected
by a relative error of the order of $\sim{a^\star}^2$. 

In Table~\ref{tableresults} and in Fig.~\ref{discrepancy}, we show the
relative difference  between the frequencies computed according to GR and to 
the EDGB theory, 
\begin{equation}
\epsilon_i=\frac{\nu_i^{\tn{EDGB}}-\nu_i^{\tn{GR}}}{\nu_i^{\tn{GR}}}\qquad
i=\varphi, r, \theta,\tn{nod}\,,\tn{per}\,,
\end{equation}
for different
values of $\alpha/M^2$ and ${a^\star}$. We assume a fiducial value of the BH
mass $M=5.3\,M_\odot$, corresponding to the estimated mass of GRO
J1655-40  \citep{Motta:2013wga}, and an emission radius
$r=1.1\,r_{\tn{ISCO}}$.  The frequencies $\nu_i^{\tn{GR}}$ are those
obtained by using the Kerr metric, and are given by
\begin{eqnarray}
\nu_\varphi^{\tn{GR}}&=&\frac{1}{2\pi}\frac{M^{1/2}}{r^{3/2}+ {a^\star}
M^{3/2}}\,, \label{nu1GR}\\
\nu_r^{\tn{GR}}&=&\nu_\varphi^{\tn{GR}}\left(1-\frac{6M}{r}+
8{a^\star}\frac{ M^{3/2}}{r^{3/2}}- 3{a^\star}^2\frac{
M^2}{r^2}\right)^{1/2}\,,\label{nu2GR}\\
\nu_\theta^{\tn{GR}}&=&\nu_\varphi^{\tn{GR}}\left(1-4{a^\star}\frac{
M^{3/2}}{r^{3/2}}+ 3{a^\star}^2\frac{ M^2}{r^2}\right)^{1/2}\,.
\label{nu3GR} \end{eqnarray}
  In the following, consistent with the slow-rotation approximation, we will consider
  Eqns.~(\ref{nu1GR})-(\ref{nu3GR}) expanded to first order in $a^\star$.

As shown in Table~\ref{tableresults} and in Fig.~\ref{discrepancy}, the
deviations increase with $\alpha/M^2$, and also with the BH spin ${a^\star}$.
They reach values as high as $\sim 1-4$\%; only $\epsilon_\tn{nod}$ can be as high
as $\sim 13$\%.
We also find that each $\nu_i^{\tn{GR}}$ is always
smaller than the corresponding $\nu_i^{\tn{EDGB}}$, thus all
$\epsilon_i$ are greater than zero.
We note that since we neglect $O(a^{\star\,2})$ terms, the
$\alpha/M^2\to0$ limit of our results differs from GR 
by terms of the same order, with the exception of $\nu_\tn{nod}$, which is proportional to 
$a^\star$ and therefore has a relative error of $\epsilon_\tn{nod}\propto a^\star$.
  This should be considered as an order-of-magnitude estimate of the errors, since we do not know the 
  $O((a^\star)^2)$ terms in the frequencies $\nu_i^{\tn{EDGB}}$ when $\alpha\neq0$.
\subsection{Testing gravity with {\it LOFT}}\label{sec:analysis}
\begin{figure*}[htbp]
\centering
\includegraphics[width=8.5cm]{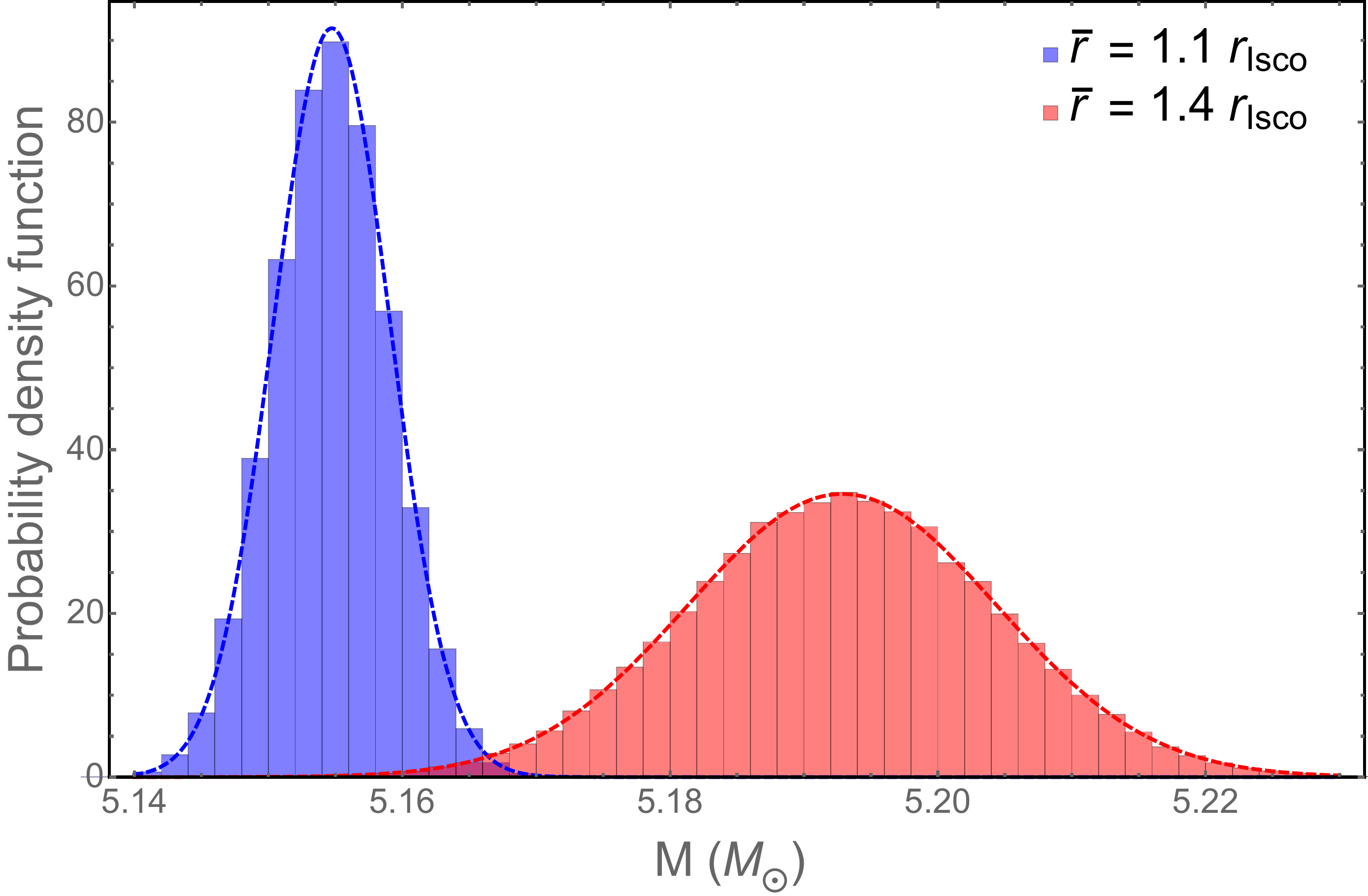}
\includegraphics[width=8.7cm]{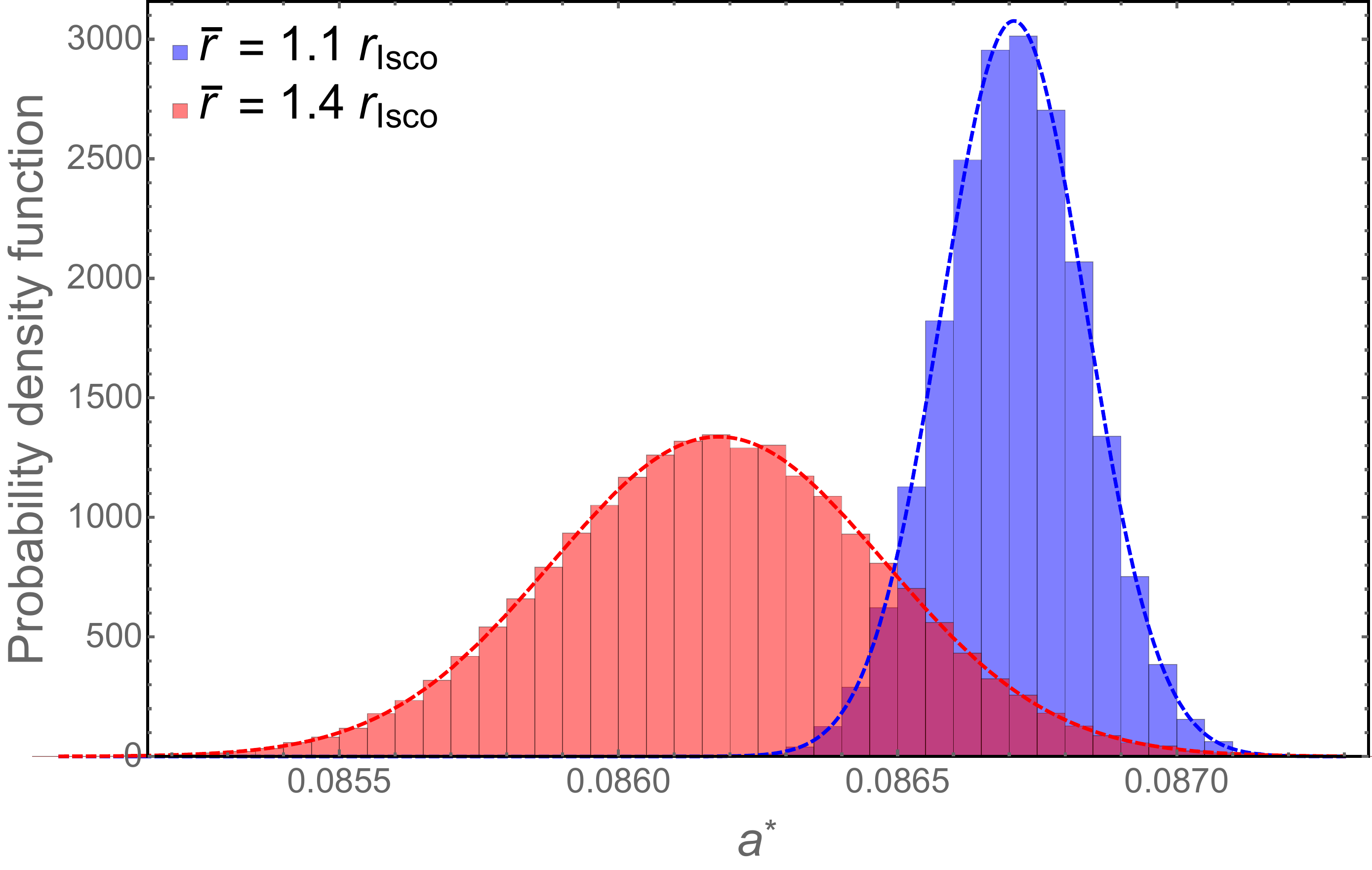}
\caption{Distribution of the BH mass $M$ (left panel) and spin $a^\star$
(right panel) as obtained from the measurement
  of the epyciclic frequencies
$(\nu_\varphi,\nu_{\tn{per}},\nu_{\tn{nod}})$ at two different radii
$r$. The generated frequencies follow Gaussian distributions with 
relative widths 15 times lower than those measured with the RXTE/PCA 
from GRO~J1655-40, as expected for the {\it LOFT}-LAD instrument.
A BH mass
 $\bar{M}=5.3 M_\odot$ and spin parameter $\bar{a}^\star=0.1$ are 
 adopted here in EDGB gravity with the maximal
coupling constant ($\alpha/\bar{M}^2=0.691$).}
\label{1danalysis}
\end{figure*}

According to the RPM, the three simultaneous QPO frequencies
$(\nu_\varphi,\nu_{\tn{per}},\nu_{\tn{nod}})$ are 
all generated at the same radial coordinate in the accretion flow.
For each such triplet,
the GR Eqns.~(\ref{nu1GR})--(\ref{nu3GR}) provide a system of three
equations for the three unknown parameters  $(M,{a^\star},r)$, which can
be solved analytically. In the EDGB modified gravity, however, the coupling 
parameter $\alpha/M^2$ which measures deviations from GR enters as an extra 
unknown parameter. In general, at least one more QPO triplet would be required 
in the RPM in order also to derive the 
$\alpha/M^2$ parameter\footnote{In principle, if an independent, very precise measurement 
of the BH mass $M$ were available, then a single triplet would suffice. The precision
would have to be 1\% or better, see Fig.~\ref{1danalysis} (left panel). This 
compares unfavorably with the precision afforded by present optical/NIR mass measurements,
see e.g. \citet{Ozel:2010wx} and references therein.}.

As we noted in Sec.~\ref{sec:RPM},  it is expected that the 
{\it LOFT} very high 
effective area will allow for the detection of  
multiple QPO triplets corresponding to different 
radii in individual stellar mass BHs.

In this section, we explore the potential of such observations for testing GR in a 
strong-field/high-curvature regime and discriminate GR against EDGB gravity. 

First, we concentrate on the case in which two different QPO
triplets are measured. We proceed as follows. \begin{itemize}
\item We fix the BH mass $\bar{M}=5.3~M_\odot$, and choose three values
for the BH spin ($\bar{a}^\star=0.05, 0.1,0.2$). We then choose values of 
$\alpha/\bar{M}^2$ consistent with the theoretical bound~(\ref{BHcond2}).
\item 
Using the EDGB equations, we generate two sets of frequencies
$\nu_{\tn{ref1}}=(\nu_\varphi,\nu_{\tn{per}},\nu_{\tn{nod}})_1$ and
$\nu_{\tn{ref2}}=(\nu_\varphi,\nu_{\tn{per}},\nu_{\tn{nod}})_2$
corresponding to two different emission radii
$r_1/r_{\tn{ISCO}}=1.1$ and $r_2/r_{\tn{ISCO}}=1.4$, respectively.  
As discussed in the Introduction, we
assume that these are the QPO frequencies measured by {\it LOFT}, with 
corresponding  uncertainties,  
$(\sigma_\varphi,\sigma_{\tn{per}},\sigma_{\tn{nod}})$,
$15$ times smaller than those measured with  
the {\it RXTE}-PCA from GRO~J1655-40 (see Eq.~(\ref{freq})).
\item
We then interpret these simulated data as if they were generated around a Kerr BH,
and solve Eqns.~(\ref{nu1GR})--(\ref{nu3GR}) to infer
the values of $(M_j,a^\star_j,r_j)$, $j=1,2$ corresponding 
to the two QPO triplets.  If the triplets
$\nu_{\tn{ref1}}, \nu_{\tn{ref2}}$ were
generated in GR 
(i.e. with $\alpha/\bar{M}^2=0$), then this procedure would yield 
the same values of the mass and spin parameters ($M_1=M_2$ and 
$a^\star_1=a^\star_2$), to within statistical and numerical 
uncertainties,  and  
truncation error due to neglecting $O({a^\star}^2)$ terms.  Conversely,
when $\alpha/\bar{M}^2\neq0$, it can be expected that different values 
of $M_1\neq M_2$ and
$a^\star_1\neq a^\star_2$ are derived.
\end{itemize}

To quantify this discrepancy, we use a Monte Carlo approach. 
For the selected values of $\bar{M}, \bar{a}^\star, \alpha/\bar{M}^2$, 
we consider $2\times (N=10^5)$ triplets
$(\nu_\varphi,\nu_{\tn{per}},\nu_{\tn{nod}})_{j=1,2}$,
with a Gaussian distribution,
centered around the values $\nu_{\tn{ref1}}$ and $\nu_{\tn{ref2}}$, 
(computed with the EDGB theory)
with standard deviation $(\sigma_\varphi,\sigma_{\tn{per}},\sigma_{\tn{nod}})$. 
Then, we use Eqns.~(\ref{nu1GR})--(\ref{nu3GR}) (which assume GR).  In Figure~\ref{1danalysis} we
show the distribution of mass (left panel) and spin (right panel)
obtained using this procedure, assuming  $\bar{M}=5.3\,M_\odot$,
$\bar{a}^\star=0.10$,  and $\alpha/\bar{M}^2=0.691$. 
\begin{figure*}[h]
\centering
\includegraphics[width=17cm]{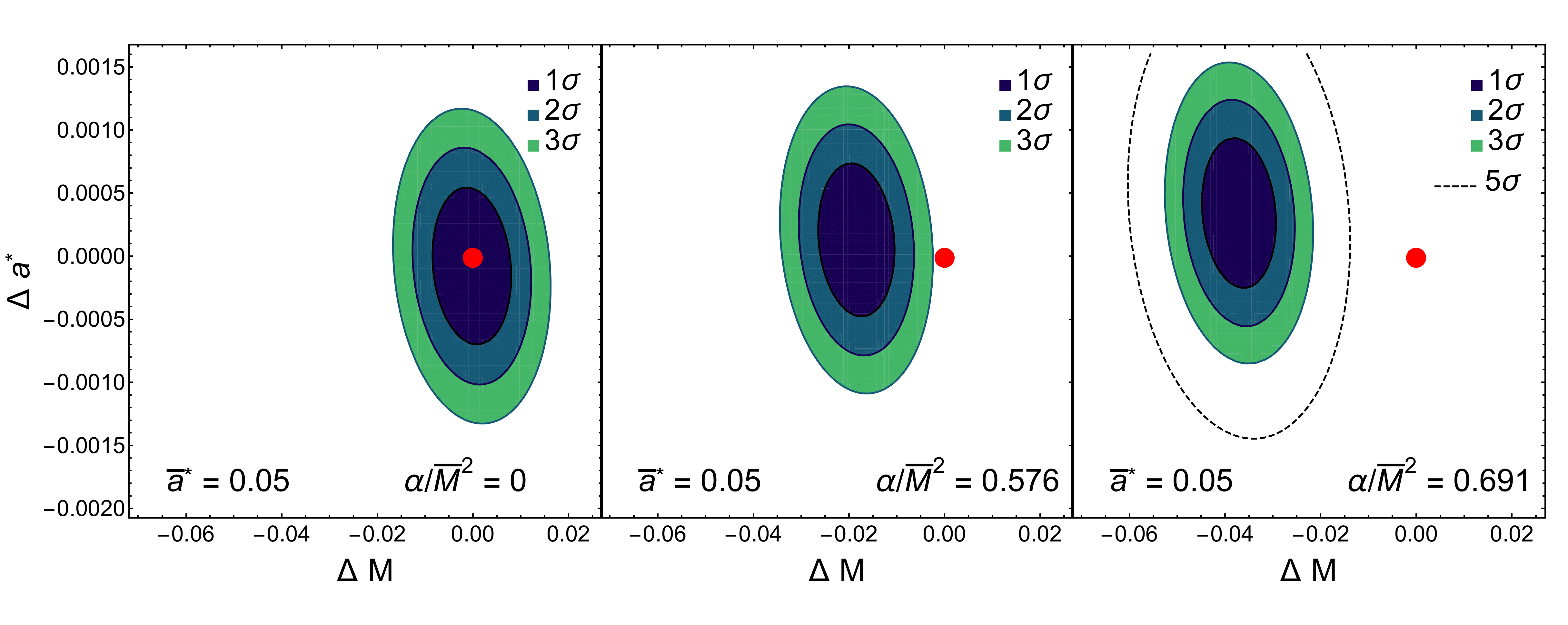}
\caption{Confidence levels  with which GR can be tested against the EDGB theory (see
text) are plotted in the $(\Delta M,\Delta a^\star)$ plane. The red dot
is the origin of the plane. 
The parameters corresponding to the two injected sets of frequencies
$(\nu_\varphi,\nu_{\tn{per}},\nu_{\tn{nod}})_j$ are
$\bar{r}_j/r_{\tn{ISCO}}=1.1,1.4$, $\bar{M}=5.3\,M_\odot$, and $\bar{a}^\star=0.05$. 
Remember that these frequencies are computed using the EDGB theory.
In the left panel, we assume that the coupling constant of
the EDGB theory, $\alpha/\bar{M}^2$, is zero, so that the EDGB theory
coincides with
GR. The red dot falls in the center of the $1\sigma$ ellipse, showing
that the the two sets of frequencies are fully compatible with GR and
that our statistical analysis is correct.
In the right panel, we set
$\alpha/\bar{M}^2=0.691$, i.e. the maximum value allowed by the EDGB theory.
The red dot is well outside the 
$5\sigma$ confidence level (dashed line), showing that in this case
we would be able to exclude that the two sets of frequencies are
generated according to GR.
In the middle panel $\alpha/\bar{M}^2=0.576$ and the red dot touches the
$3\sigma$ confidence level, i.e. the frequencies would be incompatible
with GR to this confidence level.  
}
\label{2danalysis_pre}
\end{figure*}
\begin{figure*}[htbp]
\centering
\includegraphics[width=12.5cm]{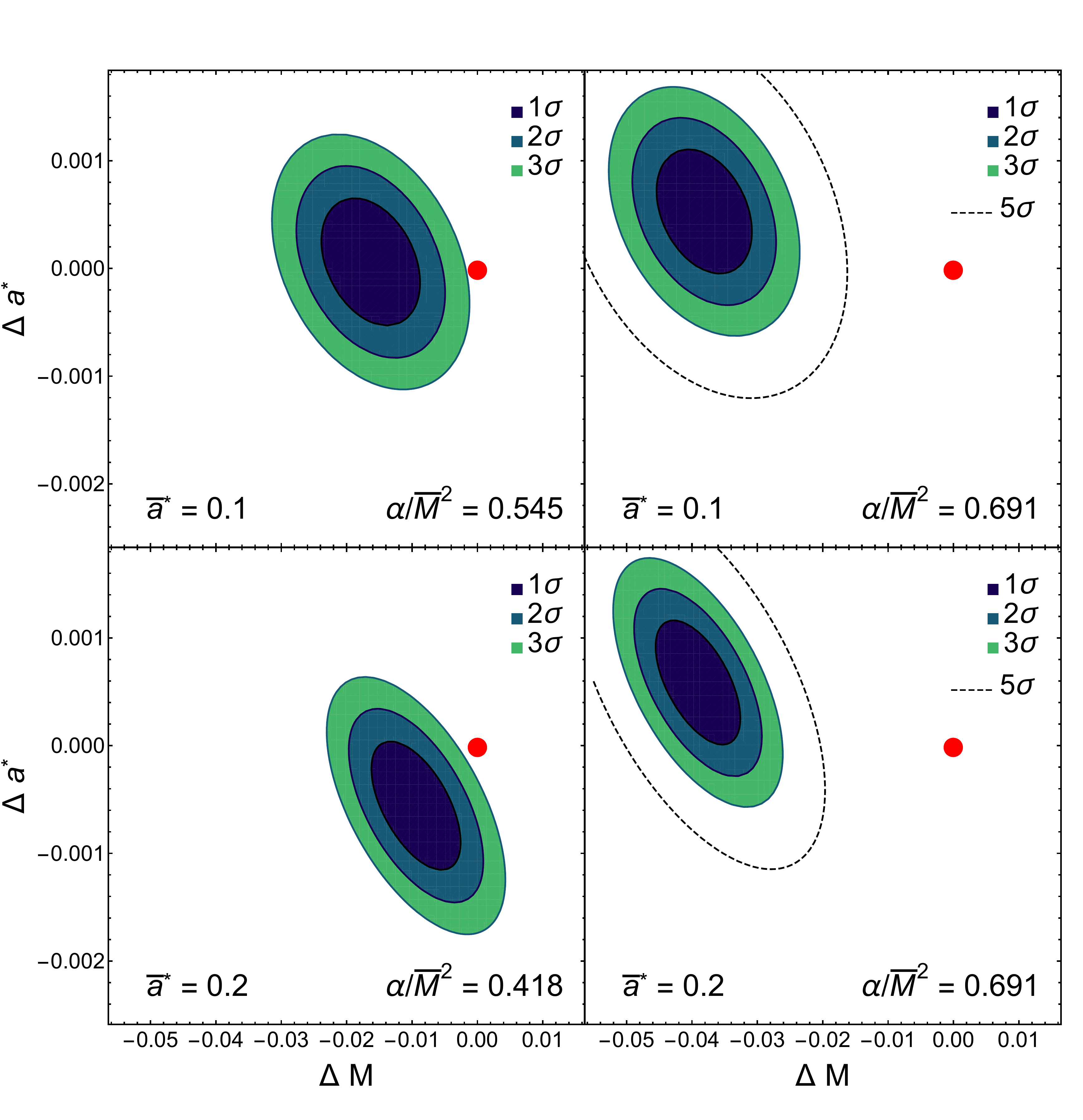}
\caption{Same as Fig.~\ref{2danalysis}, with $\bar{a}^\star=0.1,0.2$ (upper
and lower panel, respectively). $\alpha/\bar{M}^2$ has the maximum value in the
right panels, and the threshold value for which GR would be excluded at
$3\sigma$ confidence level in the left panels.}
\label{2danalysis} \end{figure*}
In order to establish to what extent these distributions are
compatible/incompatible with $M_1=M_2$ and $a^\star_1=a^\star_2$, i.e.,
with GR, we perform the following analysis.

Let us define 
\[
\Delta M=M_1-M_2, \quad \Delta a^{\star}=a^\star_1-a^\star_2, \quad
\Delta r=r_1-r_2~.
\]
We verified that the distribution of the variables
$\vec{\mu}=(\Delta M, \Delta a^{\star},\Delta
r)$ is consistent with a Gaussian distribution
$\mathcal{N}(\vec{\mu},{\Sigma}={\Sigma}_1+{\Sigma}_2)$
with zero expectation value.
We now construct the $\chi^2$ distributed variable with 3 degrees of freedom:
\begin{equation}
\chi^{2}=(\vec{x}-\vec{\mu})^{\tn{T}}{\Sigma}^{-1}(\vec{x}-\vec{\mu});
\end{equation}
$\chi^2=c$ defines the ranges
of $\Delta M$, $\Delta a^{\star}$, and $\Delta r$ at the confidence level specified by $c$.  We use 
$c=3.53,8.03,14.16$, which correspond to the $1\sigma$, $2\sigma$, and $3\sigma$ confidence
levels in a Gaussian distribution equivalent, respectively (i.e. a probability of 
32\%, 5\%, 0.3\%).

In the three panels of Figure~\ref{2danalysis_pre},
we show the results of this statistical analysis applied to the case
when the two simulated QPO triplets $(\nu_\varphi,\nu_{\tn{per}},\nu_{\tn{nod}})_j,
j=1,2$, are computed assuming $\bar{M}=5.3\,M_\odot$, $\bar{a}^\star=0.05$,
$\bar{r}_j/r_{\tn{ISCO}}=1.1,1.4$, and three values of  $\alpha/\bar{M}^2$, 
($\sim$ 0, 0.576, 0.691).
\begin{figure*}[ht]
\centering
\includegraphics[width=8.7cm]{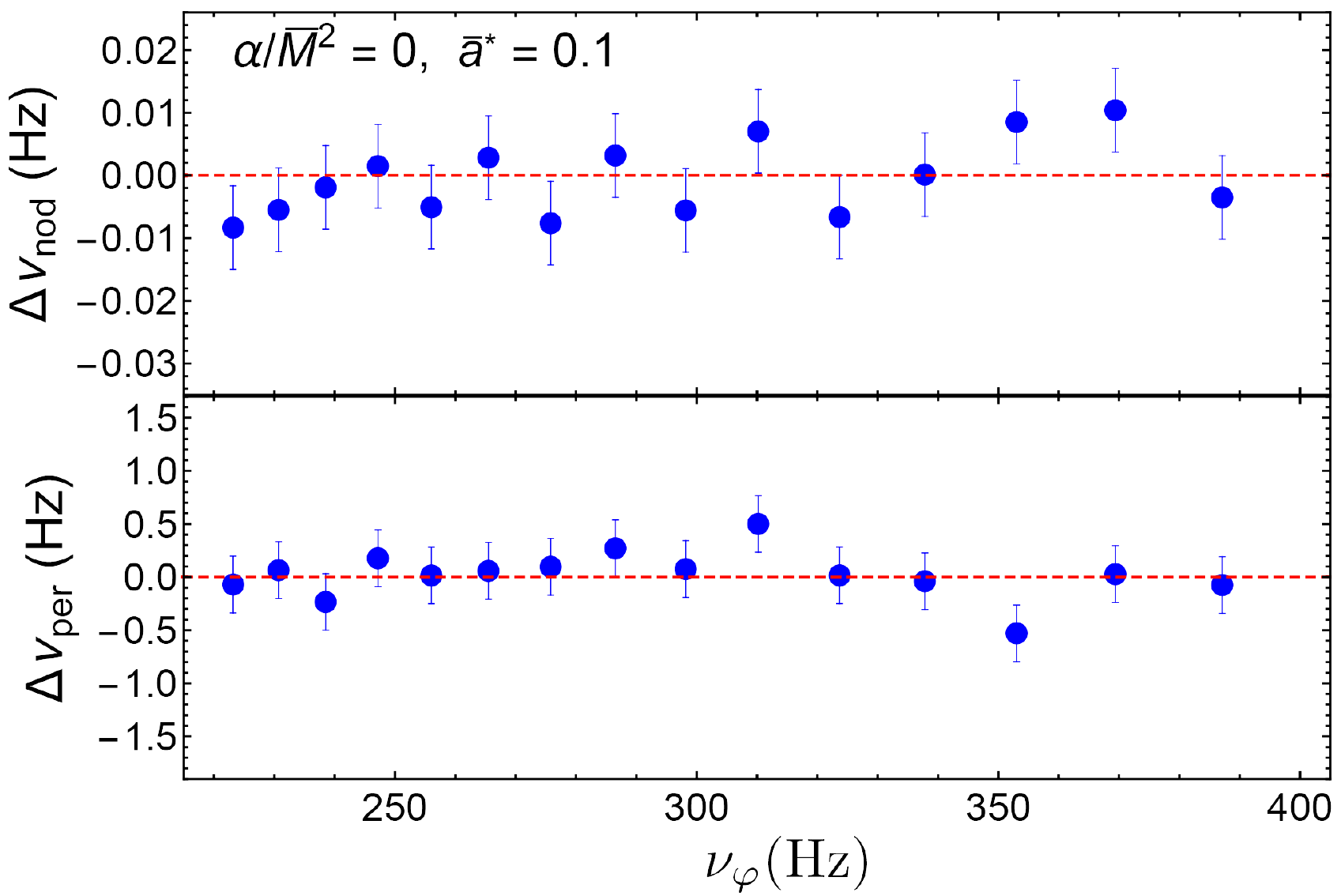}
\includegraphics[width=8.8cm]{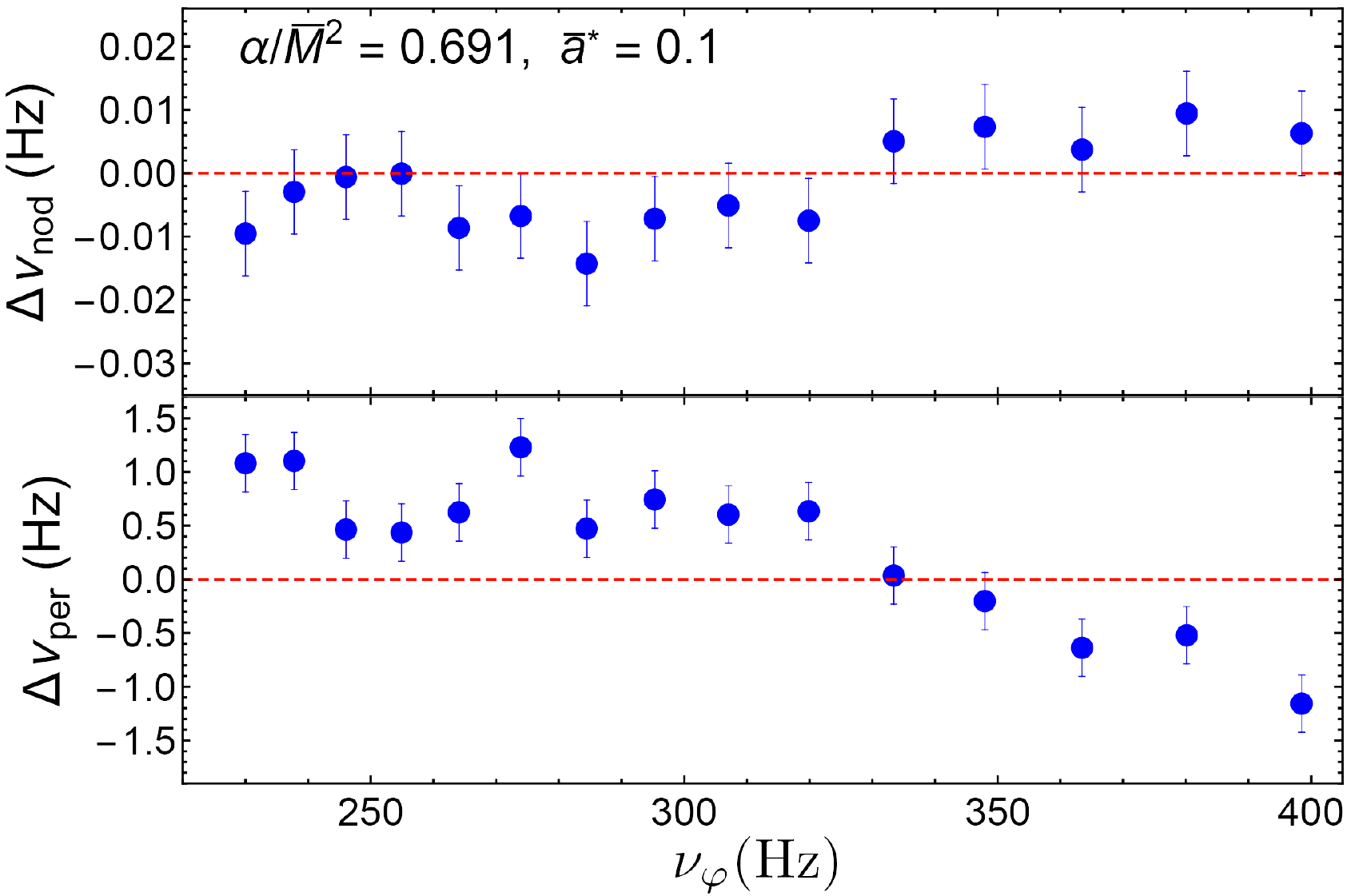}\\
\caption{$\Delta_\tn{nod}=(\nu_\tn{nod}-\hat{\nu}_\tn{nod})$
and $\Delta_\tn{per}=(\nu_\tn{per}-\hat{\nu}_\tn{per})$ are shown as
functions of the azimuthal frequency
$\nu_\varphi$, for BH configurations with $\bar{M}=5.3M_{\odot}$,
$\bar{a}^\star=0.1$ and coupling
parameter $\alpha/\bar{M}^2=(0,0.691)$. $\hat{\nu}_\tn{per,nod}$ correspond to
the periastron and
nodal frequencies computed for the best-fit parameters
$(\hat{M},\hat{a}^\star)$, which have been obtained by
minimizing the chi-square variable (\ref{chisquare}). The error bar is
$1\sigma$.}
\label{chisquarefig} \end{figure*}

Each panel shows the regions in the parameter space $(\Delta
M,\Delta a^*)$ which correspond to the $1\sigma$, $2\sigma$, and $3\sigma$
confidence levels.  The red dot corresponds to $\Delta M=0, \Delta a^*=0$.
The left panel refers to the case $\alpha/M^2 = 0 $ and the EDGB theory
coincides with GR;
the ellipses which delimit the  confidence levels are centered on the
red dots, consistent with 
two coincident values of $M$ and $a^\star$, i.e. $\Delta M=0$  and
$\Delta a^*=0$.
 
Conversely, in the right panel, the largest value of $\alpha/M^2$ allowed
by EDGB theory is used; the  red circle lies well outside the $5\sigma$
confidence ellipse, demonstrating that the two QPO triplets cannot be
interpreted within GR.
We note that this simulation is inconsistent with $\Delta M=0$, but it
is still consistent with $\Delta a^\star=0$.  This is
due to the fact that the determination of the mass is more sensitive than the
determination of the spin to the underlying gravity theory (at least for the 
slowly rotating BHs with $a^\star\le0.2$ discussed here).

Furthermore, Fig~\ref{2danalysis_pre} shows that 
for $\alpha/\bar{M}^2=0.576$ (middle panel), 
$\Delta M$ would be incompatible with 0
at $3\sigma$. Choosing $3\sigma$ 
as a threshold to assess the compatibility of GR with these
simulated data,
we conclude that when $\bar{a}^\star=0.05$, 
we would be able to discriminate between GR and EDGB theory for 
values of the coupling constant in the 
$\alpha/\bar{M}^2\in (0.576, 0.691)$ range.

In Fig~\ref{2danalysis} we show plots similar to those in 
Fig~\ref{2danalysis_pre}
for larger values of the BH angular momentum, i.e., 
$\bar{a}^\star=0.1$ (top panels) and $\bar{a}^\star=0.2$ (bottom panels). 
We adopt $\alpha/\bar{M}^2=0.691$ in the right panels and the
values of $\alpha/M^2$ corresponding to the $3\sigma$ threshold,
$\alpha/\bar{M}^2=0.545$ for $\bar{a}^\star=0.1$ and $\alpha/\bar{M}^2=0.418$
for $\bar{a}^\star=0.2$ in the left panels. We see
that as the BH spin increases, 
regions of lower $\alpha/M^2$ can be explored using this method.
Therefore, we expect that BHs spinning faster than 
$a^\star=0.2$ will allow us to test GR
against the EDGB theory for smaller  values of the
coupling constant\footnote{It should be
noted that our results for $a^\star=0.2$ are to be considered as an
extrapolation since, as mentioned above, we neglected terms of the 
order of $O(a^\star)^2$; in a subsequent paper, we will generalize the
EDGB BH solution and extend our analysis to consistently include
second-order terms in the BH spin.}.

Let us now discuss the case in which
multiple ($> 2$) QPO triplets are detected from the same BH.
We adopt the approach of considering two of the QPO 
frequencies of each triplet as a function of the third frequency 
(see Fig.~\ref{chisquarefig}), 
as commonly done in observational studies of QPOs \citep{Motta:2013wga}.
We adopt the following procedure.

\begin{enumerate}
\item 
We fix $\bar{M}=5.3M_{\odot}$ and
$\bar{a}^\star=0.1$ and  choose 15  equally spaced values of $r$ 
in the range $r=\left[1.1- 1.6\right]r_{\tn{ISCO}}$.
We then compute two sets of 15 QPO triplets 
$(\nu_\varphi,\nu_\tn{nod},\nu_\tn{per})_{j=1,\dots,15}$ corresponding to 
$\alpha/\bar{M}^2=0$ and $\alpha/\bar{M}^2=0.691$.

\item We simulate the QPO frequency triplets by drawing  
from Gaussian distributions, centered at the frequency values computed
in point 1, with the standard deviations  expect from observation with the 
{\it LOFT}-LAD (see Sec. \ref{sec:RPM}).
\item Using Eq.~(\ref{nu1GR}), we express the radial coordinate as
function of $\nu_{\varphi}$, and the BH mass and spin,
$r=r(\nu_\varphi,M,a^\star)$.  Replacing this expression into
Eqns.~(\ref{nu2GR})-(\ref{nu3GR}), we obtain the functions
$\nu^\tn{GR}_\tn{nod}(\nu_\varphi,M,a^\star)$ and
$\nu^\tn{GR}_\tn{per}(\nu_\varphi,M,a^\star)$.  
\item For each drawn value of $\nu_\varphi$ (see point 2),
we span the $M-a^\star$ parameter space (with $4~M_\odot \leq M \leq
10~M_\odot$ and $0.01\leq a^\star\leq 0.2$) and minimize the
chi-square variable 
\begin{equation}\label{chisquare}
\chi^2=\sum_{j=1}^{15}\left[\frac{(\nu_\tn{nod}-\nu^\tn{GR}_{\tn{nod}})_{j}^2}{\sigma^2_\tn{nod}}
+\frac{(\nu_\tn{per}-\nu^\tn{GR}_{\tn{per}})_{j}^2}{\sigma^2_\tn{per}}\right]\,
\end{equation} 
where $\nu_\tn{nod}$ and $\nu_\tn{per}$ are those drawn at point 2. 
This is equivalent to fitting the dependence of the simulated values 
$\nu_\tn{nod}$ and $\nu_\tn{per}$ on $\nu_\varphi$ based on GR alone. 
The $\chi^2$  has 
average $E[\chi^2]=2\times 15-2=28$ and 
standard deviation $\sigma[\chi^2]=\sqrt{2\times E[\chi^2]}\simeq 7.5$.  
\end{enumerate} 
The minimum of Eq.~(\ref{chisquare}) corresponds to the best-fit
parameters $(\hat{M},\hat{a}^\star)$. For $\alpha/\bar{M}^2=0$ we obtain
$\chi^2\simeq22$, which is within 1$\sigma$ of the expected value:
this indicates that the observed data agree with the theoretical model
based on GR predictions.  Conversely, for EDGB gravity with the maximal
coupling  $\alpha/\bar{M}^2=0.691$, the minimization yields $\chi^2\simeq134$,
which is incompatible with $E[\chi^2]=26$ at more than $14\sigma$. 

The results of this analysis are shown in Fig.~\ref{chisquarefig}, where we plot the
differences $\Delta\nu_\tn{nod}=(\nu_\tn{nod}-\hat{\nu}_\tn{nod})$ and
$\Delta\nu_\tn{per}=(\nu_\tn{per}-\hat{\nu}_\tn{per})$, with
$\hat{\nu}_\tn{per/nod}=\nu^\tn{GR}_\tn{per/nod}(\nu_\varphi,\hat{M},\hat{a}^\star)$.  
The left panel shows that for $\alpha/\bar{M}^2=0$ both $\Delta\nu_\tn{nod}$ and
$\Delta\nu_\tn{per}$ fluctuate around zero 
to within $\sim 1 \sigma$.
For $\alpha/\bar{M}^2=0.691$ (right panel) 
a clear trend is apparent, especially in $\Delta\nu_\tn{per}$, 
as a function of the azimuthal frequency $\nu_\varphi$. 
\section{Concluding Remarks}\label{sec:concl}
In this article, we have shown that the QPO frequencies can be used to test 
GR against alternative theories of gravity, in the strong-field/high-curvature regime,  
and to derive constraints on the parameters which characterize these
theories. 
In particular, we have considered one of the best motivated alternatives
to GR, i.e. the EDGB theory, 
and we have adopted for the QPOs the Relativistic Precession Model.
We have shown that the high sensitivity of the proposed ESA M-class
mission {\it LOFT}
can provide stringent constraints on the parameter $\alpha/M^2$ 
which characterizes this theory.
These constraints would be stronger than current observational bounds coming from the orbital decay rate of low-mass
X-ray binaries (\citet{Yagi:2012gp}, see Sec.~\ref{sec:slBH}) by a factor $\sim4-5$.

Our analysis was carried out considering 
slowly rotating  BHs ($a^\star\lesssim0.2$).
Since more stringent constraints on  $\alpha/M^2$ can be found for larger 
values of the black hole spin, we plan to extend this work to include 
higher-order spin corrections.
Moreover, in a future study, we will consider alternative QPO models.


\section*{Acknowledgements}
L.S. acknowledges partial support from PRIN-INAF 2011.
P.P. acknowledges support from FCT-Portugal through project IF/00293/2013.
\vskip 2cm


\bibliography{bibnote}
\bibliographystyle{apj}

\end{document}